# Multi-Placement Structures for Fast and Optimized Placement in Analog Circuit Synthesis

Raoul F. Badaoui and Ranga Vemuri
{rbadaoui,ranga}@ececs.uc.edu

## Abstract

*This paper presents the novel idea of multi-placement structures, for a fast and optimized placement instantiation in analog circuit synthesis. These structures need to be generated only once for a specific circuit topology. When used in synthesis, these pre-generated structures instantiate various layout floorplans for various sizes and parameters of a circuit. Unlike procedural layout generators, they enable fast placement of circuits while keeping the quality of the placements at a high level during a synthesis process. The fast placement is a result of high speed instantiation resulting from the efficiency of the multi-placement structure. The good quality of placements derive from the extensive and intelligent search process that is used to build the multi-placement structure. The target benchmarks of these structures are analog circuits in the vicinity of 25 modules . An algorithm for the generation of such multi-placement structures is presented. Experimental results show placement execution times with an average of a few milliseconds making them usable during layout-aware synthesis for optimized placements.*

## 1. Introduction

In the design of integrated circuits, placement is a major step that sets the coordinates of the various blocks present in the circuit on a layout surface. The synthesis process of analog circuits uses layout generation information within its sizing search loop for accuracy in performance estimation. A survey of placement algorithms [4] shows the major directions in analog placement approaches: optimization based and template based.

**Optimization based** techniques use heuristic algorithms on sized circuits using methods such as simulated annealing and genetic algorithms to meet the specified performance constraints. The KOAN/ANAGRAM [2] optimization-based placement tool belongs to this class. Other research includes Zhang's [3] work using genetic algorithms and Gielen's LAYLA [6].This method yields good placements optimizing interconnect wire-lengths. Its major drawback is convergence time which makes it hard to use in a layout-inclusive sizing process.

**Template based** techniques try to reduce the time search algorithms take in optimization-based techniques so that they can be used in a layout-inclusive sizing process. Most of these works rely on procedural module generators to describe layout templates such as BALLISTIC [1], MOGLAN [8, 7] or MSL [5]. Expert knowledge is used to design a layout template for an unsized circuit using a specific fixed placement of blocks. These templates take as input the sizes and other design parameters of the circuit and instantiate a layout, iteratively, during a synthesis process. Speed is the major advantage of this method. However, its drawback lies in its inability to explore possible good performance for the circuit that might exist for certain sizes if the circuit were to be placed differently than in the template.

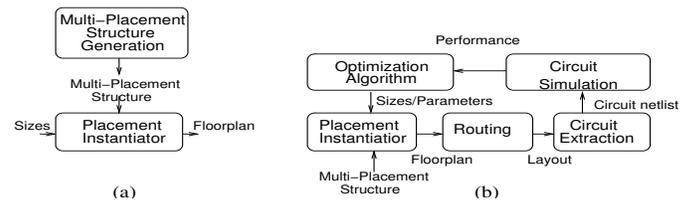

Figure 1: *(a) One-time Generation of the multi-placement structure and its (b) use in synthesis*

The proposed approach aims at retaining the benefits of both the techniques described above: a fast instantiation of layout for layout-inclusive synthesis and various placement possibilities for various input sizes (No restriction to a single, pre-defined template). It is intended for sizing analog circuits of complexity ranging up to 25 modules.

Our approach consists of a one-time generation of a multi-placement structure for a specific topology as shown in Figure 1.a . The obtained structure would be used in a layout-inclusive synthesis process in the following manner: It is provided with numerical sizes from an optimization tool and returns a specific floor-plan for the circuit. The proposed synthesis loop is shown in Figure 1.b . For different sizes given, the aim is to have the best floor-plan returned.

The rest of the paper is organized as follows: Section 2 defines the multi-placement structure and how it handles a sampled coverage of the sizing search space. Section 3 describes one algorithm to generate that multi-placement structure. Finally, Section 4 presents experimental results to support the feasibility and effectiveness of our method.

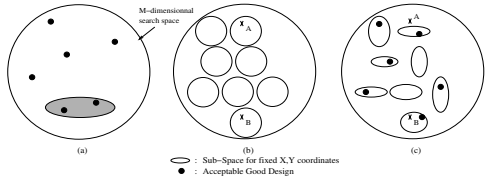

Figure 2: *Representation of a M-dimensional search space*

## 2. Multi-Placement Structure

### 2.1. Definition of a Multi-Placement Structure

A multi-placement structure is generated once for a specific circuit topology. As shown in figure 1.b, this structure can be used iteratively in a synthesis process. It instantiates the most suitable placement of blocks corresponding to the sizes and parameters fed to the placement instantiator. This section defines the various functions and structures comprising the multi-placement structure.

A circuit is defined as a set of $N$ blocks while a block is any module defined by its module generator functions. The variables $w_i$ and $h_i$ represent the width and height of block $i$ while constants $w_{m_i}$, $h_{m_i}$, $w_{M_i}$ and $h_{M_i}$ are set by the designer as the minimum and maximum widths and heights of block $i$.

A specific placement of the set $B$ of blocks would be defined as a set of $x_i$ and $y_i$ values representing the coordinates of blocks on the floor-plan.

At the beginning of our multi-placement structure generation process, we have a circuit with the widths, heights, $x$ and $y$ coordinates as unknown and variable values.

The aim is to generate a structure that maps each set of $w_i$(widths) and $h_i$(heights) of all blocks to a set of $x_i$ and $y_i$ coordinates representing the best placement to use for the specified widths and heights. That structure would mathematically be represented as the function $\mathcal{M}$, with $V = \bigcup_{i=0}^{N}(w_i, h_i)$:

$$\begin{aligned} \mathcal{M}(V) : \quad \mathcal{N}^{2N} &\mapsto \Pi \\ V &\mapsto p, p \in \Pi \end{aligned} \quad (1)$$

Set $\Pi$ would represent the set of placements stored in a multi-placement structure while vector $V$ and its $w_i$ and $h_i$ values consist of the possible dimensions of the various blocks. The resulting placement $p$ would be the best placement to use for those specific widths and heights of the blocks.

Hence, if such a function is built, it can be used during circuit synthesis as follows: First translate the proposed device sizes into widths and heights of the modules using module generator functions, then use the function $\mathcal{M}$ to obtain the placement that would best suit the proposed device sizes.

To illustrate multi-placement structures, a M-dimensional search space for some arbitrary circuit is represented as a two-dimensional circle in Figure 2.a. M is equal to the number of parameters in the circuit added to the (x,y) coordinates variables of the blocks. The black dots in the figure represent potential good solutions of the design problem. When using templates to generate the layouts, the placement is set to a fixed set of (x,y) coordinates. The sizing algorithm is hence constrained to a sub-space of the M-dimensional search space. In Figure 2.a, a shaded elliptic area illustrates conceptually what such a constraint imposes on the sizing algorithm. As shown, numerous *good* solutions are hidden in the non-shaded area. Thus, the synthesis process is not able to explore solutions outside its grey shaded area and find potential sub-optimal solutions. An exhaustive search of the whole M-dimensional space is practically impossible for time constraints. The proposed idea tries to include most possible good solutions of the search space in a reduced search space of the synthesis algorithm.

A fixed placement with specific (x,y) coordinates for the blocks is represented as one grey shaded area, thus, a multi-placement structure and its set of placements $\Pi$ would be represented as a set of grey shaded areas such as the ones shown in Figure 2.b. These elliptic areas can be overlapping in the search subspace of the synthesis tool ( The synthesis tool does not include (x,y) coordinates in its search parameters ). For example, points A and B represent one solution with the same values for all the parameters of the circuit. The only difference is in the (x,y) coordinates values of the blocks. Based on the definition of the Multi-Placement Structure, the latter should return one specific placement for each unique set of circuit parameter values: the best one. In Figure 2.b, both points A and B are each inside one shaded area of the search space. Thus, to make them comply with the main condition of returning only one placement, the placements stored in the Multi-Placement Structure should be shrunk in the circuit parameters range search space as it is conceptually shown in Figure 2.c. Each placement $p_j$ along with its reduced widths and heights space shall be mathematically formulated as:

$$\begin{aligned} (\Pi \times B) &\mapsto \mathcal{N}^4 \\ (p_j, B_i) &\mapsto \{w_{start_{i,j}}, w_{end_{i,j}}, h_{start_{i,j}}, h_{end_{i,j}}\} \end{aligned} \quad (2)$$

$B$ is the set of all blocks $B_i$ of the circuit. $\Pi$ is the set of all placements $p_j$ stored in the multi-placement structure. The values $w_{start}$ and $w_{end}$ ( $h_{start}$ and $h_{end}$ ) for a specific placement $p_j$ represent an interval of all possible values of $w_i$ ($h_i$) for each block $B_i$ reducing the *coverage* placement $p_j$ has. Thus, placement $p_j$ becomes valid if and only if $w_i$ ($h_i$) of every block $B_i$ lies within $[w_{start_{i,j}}, w_{end_{i,j}}]$ ( $[h_{start_{i,j}}, h_{end_{i,j}}]$). These 'start' and 'end' values should then be set during the generation of the multi-placement structure in a way to ensure that placements are only valid and used within a range of $w_i$ and $h_i$ that would make the combination of the widths, heights, $x$ and $y$ coordinates best for the performance of the circuit.

The Multi-Placement Structure shall have a structure such as the one represented conceptually in Figure 2.c : Non-overlapping placements of the circuit stored in the structure, only one of those placements returned for every set of circuit parameters fed to the structure and most good design solutions points included in its search space. The structure shall be generated once for a circuit and then used repetitively in synthesis, allowing a better convergence during synthesis, yielding a multitude of placement possibilities with a fast instantiation time suitable for synthesis. Section 2.2 defines the data structures involved in implementing the multi-placement structures defined above.

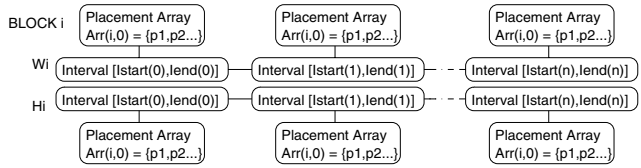

Figure 3: *Row representing a block in a multi-placement structure object*

## 2.2. Computational Implementation of the Multi-Placement Structure

Function $\mathcal{M}$ can be implemented as partially shown in Figure 3. Each block is represented by one structure as the one shown.

**2.2.1. Block-Row in a Multi-Placement Structure** When an input vector $V$, is fed to that structure, each $(w_i,h_i)$ pair is fed to the structure corresponding to Block $i$. Those two values run through a linked list of interval objects. Each interval object represents an integer interval $n$: $I_{start_{i,n}}$ to $I_{end_{i,n}}$. A linked list of such intervals has the constraint of being *ascending* and *non-overlapping*. The value fed to the linked list must fall in one of those intervals.

When a $w_i$ or a $h_i$ value finds its interval, it returns an array of numbers $Arr_{i,n}$ attached to that specific interval as shown in the figure. This array of numbers represents the indices of all placements $pj$ in which $w_i$ ($h_i$) of vector $V$ lie within $w_{start_{i,j}}$ and $w_{end_{i,j}}$ ($h_{start_{i,j}}$ and $h_{end_{i,j}}$) of that placement $p_j$. Each of these rows can be mathematically represented as the following function, $a$ being an integer:

$$\mathcal{W}_i(a)(or \mathcal{H}_i(a)) : \mathcal{N} \mapsto \Pi \atop a \mapsto \pi \subset \Pi \qquad (3)$$

The $\pi$ sets returned from the $\mathcal{W}$ and $\mathcal{H}$ functions are subsets of the $\Pi$ set. The latter is the set of all placements stored in the multi-placement structure. Every $\pi$ returned from a row $i$ of the multi-placement structure consists of a set of placements suitable to use for the $w_i$ and $h_i$ values of the input $V$ vector. $\pi$ is the equivalent of the $Arr_{i,n}$ type objects shown in figure 3. The intersection of those returned subsets should be one and only one placement, the one to use for those specified input widths and heights.

**2.2.2. The Complete Multi-Placement Structure** When all the $(w_i,h_i)$ pairs are fed to their corresponding rows in the structure, we would obtain a set of array of numbers returned from each row in the structure. The intersection of all those arrays of placements should correspond to the resulting placement from our multi-placement structure. That structure's function $\mathcal{M}$ would then simply be computed as:

$$\mathcal{M}(V) : \mathcal{N}^{2N} \mapsto \Pi \qquad (4)$$
$$V \mapsto \pi = \bigcap_{i=0}^{N}(\mathcal{W}_i(w_i) \cap \mathcal{H}_i(h_i))$$

Our goal is to have only one placement returned from function $\mathcal{M}$, thus:

$$|\mathcal{M}(V)| = 1 \qquad (5)$$

Section 3 presents how a novel algorithm is used to generate the multi-placement structure described and the way it ensures that equation 5 is observed.

## 3. The Multi-Placement algorithm

This section presents the algorithm used to build the multi-placement structure described in section 2. The goal is to generate the structures of figure 3 ensuring the realization of equation 4 and the compliance with equation 5.

Figure 4 shows the main steps and modules involved in the generation of the multi-placement structure. The two major parts of the algorithm are the **Placement Explorer** and the **Block Dimensions-Interval Optimizer**.

The **Placement Explorer** is a search-like tool that intelligently chooses various placements by selecting values for the $(x_i,y_i)$ coordinates. It then finds out which range of values of $w$'s and $h$'s yields best performance for those specific $(x_i,y_i)$ values, and sets the value of the $(w_{start},w_{end},h_{start},h_{end})$ 4-tuple accordingly. Finally, it stores this placement in a multi-placement structure such as the one described in section 2. The placement explorer obtains a *cost* value for this placement using the other part of the tool, the Block Dimensions-Interval Optimizer.

The **Block Dimensions-Intervals Optimizer** takes a placement with fixed $(x_i,y_i)$ values as input along with the 4-tuple $(w_{start}, w_{end}, h_{start}, h_{end})$. It runs a search algorithm (with the $w$ and $h$ dimensions of the blocks as variables) to try and reduce those $w$ and $h$ intervals around the values that result in the lowest wiring lengths and area for the circuit. This tool returns to the placement explorer the 4-tuple representing the reduced dimensions interval fed in along with an average value of the cost induced by the various wire lengths and areas encountered during the search. The best attained value of that cost is also returned. The said average value returned would be used as the cost indicator of the placement explorer as stated above. A detailed description of the algorithm used to implement the tools described above follows.

### 3.1. Placement Explorer

The several steps involved in the placement's explorer algorithm are shown in Figure 4 and described below.

**3.1.1. Placement Selector:** The placement selector follows a simulated annealing based method to perform its task. It initially selects a random placement for the topology. The $\{w_{start_{i,j}}, w_{end_{i,j}}, h_{start_{i,j}}, h_{end_{i,j}}\}$ values of the allowable ranges of widths and heights for that placement are initially set to the minimum widths and heights of the blocks. These values are then expanded in the **Placement Expansion** step presented below.

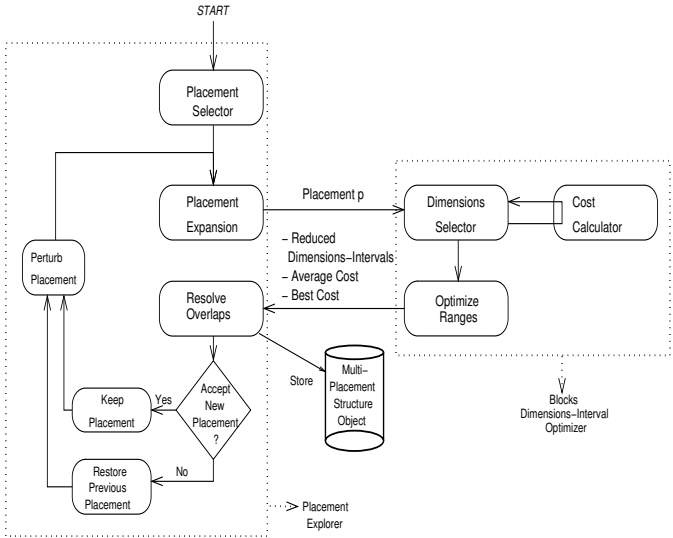

Figure 4: *Algorithm Flow Detailed Representation*

**3.1.2. Placement Expansion:** This step takes in the selected placement with its blocks' dimensions ranges set to their minimum and expands them on the floor-plan while keeping them from overlapping.

Blocks have their dimensions incremented one by one until no further expansion is possible due to overlapping or out-of-bounds constraints. This expansion would form an interval of widths and heights for the blocks.

The expanded placement $p$ is then sent to the Block Dimensions-Interval Optimizer (**BDIO**) described later in Section 3.2. This placement represents a proposed solution in the simulated annealing scheme of the placement explorer. The **BDIO** will return back to the placement explorer a cost value for this placement and a reduced range of $w$ and $h$ applicable for this placement. Those shrunk intervals are the ones that would yield best performance.

Thus, this placement should be returned by the multi-placement structure once used, if it is fed with $w$ and $h$ values that lie in the specified range.

To make the multi-placement structure comply with equation 5, and have the structure only return *one* placement for the specified $w$ and $h$ intervals, the latter interval should not intersect with any other $w$ and $h$ interval attached to any placement already stored in the multi-placement structure. Section 3.1.3 below describes the **Resolve Overlaps** step handling this issue in the algorithm

**3.1.3. Resolve Overlaps:** The $(w,h)$ variable space that is now attached to placement $p$ might intersect with an already explored placement's set of ranges. In order to ensure that equation 5 holds true, there should be no overlap between two placements' intervals of block dimensions. The **Resolve Overlaps** routine defined below has the task of resolving those overlaps and enforcing equation 5.

```
I: Set of Overlapping placements
```
**Resolve Overlaps:**
$\forall i; i : 0 \to N$
  $\forall w_i; w_i : w_{start_{i,j}} \to w_{end_{i,j}}$
  $\forall h_i; h_i : h_{start_{i,j}} \to h_{end_{i,j}}$
    $I = \mathcal{W}_i(w_i) \cap \mathcal{H}_i(h_i) \cap I$
$\forall \pi, \pi \in I$
  **Resolve Overlap**$(\pi, p_j)$

By using the $\mathcal{W}$ and $\mathcal{H}$ functions representing the multi-placement structure shown in figure 3 and initially empty, the **Resolve Overlaps** routine retrieves all the placements in the multi-placement structure that overlap with placement $p$. One by one, all these placements' overlaps are resolved using the **Resolve Overlap** routine.

The latter searches for the smallest dimension ( row ) in which the two placements are overlapping. The values of the average cost of each of the placement are then compared. The placement with a higher average cost is chosen to be shrunk in the found dimension. Shrinking a placement's interval range in one selected row consists of taking a placement row as an input along with its ($w_{start}, w_{end}, h_{start}, h_{end}$) 4-tuple vector. It either shrinks the selected $w$ or $h$ interval. If the overlapping interval to be shrunk contains completely the other placement's interval from the *start* and the *end* sides, it is forked into two placements, each assuming new shrunk intervals on each side of the un-changed placement. The shrunk placement(s) has its corresponding ($w_{start}, w_{end}, h_{start}, h_{end}$) 4-tuple values adjusted accordingly.

Having resolved overlaps, placement $p$ is then stored in the multi-placement structure using the **Store Placement** routine.

As shown below, the **Store Placement** algorithm is actually modifying the data structure represented in figure 3. It adds interval objects and splits others into two in order to keep the non-overlapping and ascending characteristics of the linked list of interval objects. The index of the placement being stored is then added to the corresponding array of indices above the placement's interval represented by one or more interval objects.

**Store Placement:**
$\forall i; i : 0 \to N$
Redefine $\mathcal{W}_i(a)$:
  $\mathcal{N} \mapsto \Pi$
  $a \mapsto \pi = \pi \cup \{p_j\}$   $\forall a, w_{start_{i,j}} \leq a \leq w_{end_{i,j}}$
Redefine $\mathcal{H}_i(a)$:
  $\mathcal{N} \mapsto \Pi$
  $a \mapsto \pi = \pi \cup \{p_j\}$   $\forall a, h_{start_{i,j}} \leq a \leq h_{end_{i,j}}$

Having resolved overlaps and stored the placement in the multi-placement structure, the algorithm follows a simulated annealing based selection criteria to choose the next placement to be evaluated. The **Accept New Algorithm** check of Figure 4 performs a condition check on the cost of the placement being explored. If placement *p* is accepted it is used to select the next placement to be evaluated through the **Perturb Placement** step of the algorithm. Otherwise, the last accepted placement is used.

**3.1.4. Perturb Placement:** The **Perturb Placement** step of the algorithm of Figure 4 uses the current accepted placement to select a new placement and sends it back to be evaluated by the placement explorer. Based on a percentage value set by the user, a set number of blocks' *x* and *y* coordinates are randomly varied. To allow some shuffling of the circuit, an out-of-bound coordinate variation is not discarded but used to shift the block back to the opposite side of the floor-plan.

As for the stopping criterion of the Simulated Annealing process, a value representing the percentage coverage of the widths and heights ranges space is calculated and updated. The placement explorer algorithm keeps running until an acceptable value (set by the user) of that percentage is reached knowing that the ideal 100% value can never be reached. The remaining uncovered percentage of the space would then be mapped to a template-like placement for backup purposes. Section 3.2 describes the algorithm used to implement the Block Dimensions-Intervals Optimizer.

## 3.2. Block Dimensions-Intervals Optimizer

The Block Dimensions-Intervals Optimizer (BDIO) is supposed to perform the following tasks on one placement handed to it by the Placement Explorer:

- Minimize the values of the intervals represented by the 4-tuple ($w_{start}$, $w_{end}$, $h_{start}$, $h_{end}$) around the values of *w* and *h* that would best produce a better performance circuit for the specific placement in question.

- Calculate an average cost for the placement considered. This average cost is the one to be returned to the Placement Explorer along with the best cost attained.

The Block Dimensions-Intervals Optimizer (BDIO) has been also implemented using a Simulated Annealing methodology. This makes our multi-placement structure generation algorithm a nested simulated annealing style algorithm. The algorithm for the **BDIO** is shown in Figure 4 on its right side. The following sections describe the tasks performed on the placement handed to the **BDIO** by the placement explorer.

**3.2.1. Dimensions Selector:** The dimensions selector selects specific values for the *w* and *h* values iteratively and sends them to a cost calculator. The proposed solution for the simulated annealing of the **BDIO** is thus a numerical value of the widths and heights of the blocks. Those values should be valid values in the range specified by the 4-tuple ($w_{start}$, $w_{end}$, $h_{start}$, $h_{end}$) of the placement being evaluated.

Having obtained a value for the cost from the cost calculator (described in Section 3.2.2), the dimensions selector perturbs the proposed *w* and *h* values by a percentage input set by the user at input.

| Circuit | Blocks | Nets | Terminals |
|---|---|---|---|
| circ01 | 4 | 4 | 12 |
| circ02 | 6 | 4 | 18 |
| circ06 | 6 | 4 | 18 |
| TwoStage Opamp | 5 | 9 | 22 |
| SingleEnded Opamp | 9 | 14 | 32 |
| Mixer | 8 | 6 | 15 |
| circ08 | 8 | 8 | 24 |
| tso-cascode | 21 | 36 | 46 |
| benchmark24 | 24 | 48 | 48 |

Table 1: *Test Benchmarks*

**3.2.2. Cost Calculator:** The cost calculator has a fixed placement along with fixed widths and heights of the blocks present in the circuit as its input. It calculates a cost for the proposed circuit based on the wire-lengths and area of that proposed design. This cost function is customizable.

At the end of the simulated annealing process whose stopping criterion is a number of iterations set by the user, the average cost value obtained is returned along with the best cost value attained. Added to that, the ($w_{start}$, $w_{end}$, $h_{start}$, $h_{end}$) 4-tuple has to be minimized around the values of *w* and *h* that have produced the best cost during the iterations. The **Optimize Ranges** step of the algorithm described below in Section 3.2.3 is used to minimize those ranges.

**3.2.3. Optimize Ranges:** The interval of the *w* and *h* intervals of the blocks represented by the ($w_{start}$, $w_{end}$, $h_{start}$, $h_{end}$) 4-tuples is optimized around the values of *w* and *h* that yielded the best cost. The values for each block *i* in the selected placement $p_j$ are adjusted in the same way the $w_{start}$ is changed below:

$$w_{start_{i,j}} = w_{best} - \frac{averagecost}{bestcost} \times (w_{end_{i,j}} - w_{start_{i,j}}) \quad (6)$$

Use of this formula makes the intervals proportional to the ratio of the average cost and best cost. The further the average cost is away from the best cost, the tighter we would like the interval to be around $w_{best_i}$ and $h_{best_i}$.

The Blocks dimensions-interval optimization algorithm returns the placement along with its reduced interval 4-tuple, its best cost and its average cost. The Placement Explorer uses the returned average cost value as a cost for its own SA-like algorithm.

Finally, the multi-placement structure would be filled with a multitude of placements, mapping to widths and heights of the blocks present in the circuit, confining with equations 1, 4, and 5. The next section will show through some example benchmarks the usage of the proposed method and its efficiency.

## 4. Experimental Results

A multi-placement structure has been generated for each of the circuits presented in Table 1. The algorithm was written in C++, and run on a SUN-Blade-1000 workstation with 2GB of RAM. Table 2 shows the details of the multi-placement structures generation. The *placements* column shows the number of possible template placements modeled in each multi-placement structure. The *instantiation* column reveals the time it takes to instantiate one placement when the structure is fed with sizes for the circuit. Those instantiation times prove to be short enough for use in a layout-inclusive synthesis process.

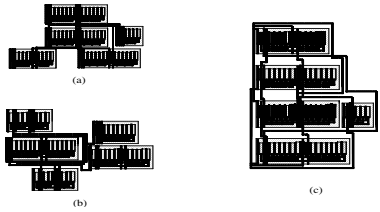

Figure 5: *Various sizes Floorplan instantiations (a,b) and template-based instantiation (c) for two-stage opamp*

| Circuit | CPU Generation Time | Placements | Instantiation |
|---|---|---|---|
| circ01 | 21m12s | 57 | **0.07s** |
| circ02 | 25m35s | 51 | **0.085s** |
| circ06 | 46m23s | 86 | **0.1s** |
| TSO | 52m45s | 82 | **0.09s** |
| SEO | 1h55m | 115 | **0.12s** |
| Mixer | 57m23s | 75 | **0.11s** |
| circ08 | 1h42m13s | 123 | **0.12s** |
| tso-cascode | 2h36m35s | 124 | **0.14s** |
| benchmark24 | 4h03m | 133 | **0.15s** |

Table 2: *Usage and Generation of the Multi-Placement Structures Generated*

The various placements generated for the various sizes provided would be optimized with respect to area and wire-length. Figures 5.a and 5.b show two instantiations of the two-stage opamp when its generated multi-placement structure is used. The template-based manual placement is shown in 5.c as a comparison. The multi-placement structure idea suits analog circuit synthesis best due to the higher need of exact layout elaboration during synthesis. Circuit *tso-cascode* is a benchmark circuit of op-amps in cascode comprised of 21 modules, comparable in size to most complex analog blocks. An instantiation of its optimized layout is shown in figure 7.

Finally, to validate the correctness of the multi-placement structure, various instantiations of the same module were conducted, following a variation in one dimension of the search space. The cost incurred from using several placements from within the multi-placement structure is shown in the top plot of Figure 6. The bottom plot shows the cost when the multi-placement structure is used; clearly, the lowest cost placement was selected, depending on the location of the proposed solution in the search space.

## 5. Conclusion

This paper presented an algorithm to generate a multi-placement structure as a way to obtain optimized placements in a synthesis loop without having to include a timely placement algorithm. The method extremely reduces the amount of time taken to generate those placements and is comparable to template-based approaches in speed. On the other hand, depending on the parameters and sizes of the circuit used, it instantiates various floor-plan settings, as if optimization-based methods were being used. It instantiates placements within milliseconds. It is applicable to most analog blocks of sizes ranging up to 25 modules. It does not require expert knowledge to pre-generate placement templates.

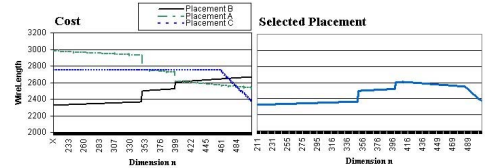

Figure 6: *Lowest Cost Selection for TwoStageOpamp Multi-Placement Structure.*

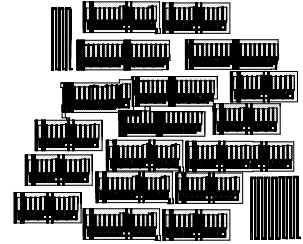

Figure 7: *Example floor-plan instantiation for circuit 'tso-cascode'*